\documentclass[12pt,preprint]{aastex}

\slugcomment{The Astrophysical Journal Letters, in press}

\shorttitle{First detection of water vapor in a pre-stellar core}
\shortauthors{Caselli et al.}

\begin{document}


\title{First detection of water vapor in a pre-stellar core}


\author{Paola Caselli\altaffilmark{1,2}, Eric Keto\altaffilmark{3}, Edwin A.  Bergin\altaffilmark{4}, Mario Tafalla\altaffilmark{5}, \\
Yuri Aikawa\altaffilmark{6}, Thomas Douglas\altaffilmark{1}, Laurent Pagani\altaffilmark{7}, Umut A.~Y{\i}ld{\i}z\altaffilmark{8}, \\ 
Floris F. S. van der Tak\altaffilmark{9,10}, C. Malcolm Walmsley\altaffilmark{2,11}, Claudio Codella\altaffilmark{2}, \\
Brunella Nisini\altaffilmark{12}, Lars E. Kristensen\altaffilmark{8}, Ewine F. van Dishoeck\altaffilmark{8,13}}


\altaffiltext{1}{School of Physics and Astronomy, University of Leeds, LS2 9JT Leeds, UK. \email{p.caselli@leeds.ac.uk}}
\altaffiltext{2}{INAF-Osservatorio Astrofisico di Arcetri, Largo E. Fermi 5, 50125 Firenze, Italy.}
\altaffiltext{3}{Harvard-Smithsonian Center for Astrophysics, 60 Garden Street, Cambridge,  MA 02138, USA.}
\altaffiltext{4}{Department of Astronomy, The University of Michigan, 500 Church Street, Ann Arbor, MI 48109-1042, USA.}
\altaffiltext{5}{Observatorio Astron\'omico Nacional (IGN), Calle Alfonso XII, 3, Madrid, Spain.}
\altaffiltext{6}{Department of Earth and Planetary Sciences, Kobe University, Nada, 657-8501 Kobe, Japan.}
\altaffiltext{7}{LERMA and UMR 8112 du CNRS, Observatoire de Paris, 61 Av. de l'Observatoire, 75014 Paris, France.}
\altaffiltext{8}{Leiden Observatory, Leiden University, PO Box 9513, 2300 RA Leiden, The Netherlands.}
\altaffiltext{9}{SRON Netherlands Institute for Space Research, PO Box 800, 9700 AV, Groningen, The Netherlands.}
\altaffiltext{10}{Kapteyn Astronomical Institute, University of Groningen, The Netherlands.}
\altaffiltext{11}{The Dublin Institute for Advanced Studies, 31 Fitzwilliam Place, Dublin 2, Republic of Ireland.}
\altaffiltext{12}{INAF - Osservatorio Astronomico di Roma, 00040 Monte Porzio Catone, Italy.}
\altaffiltext{13}{Max Planck Institut f\"{u}r Extraterrestrische Physik, Giessenbachstrasse 1, 85748 Garching, Germany.}


\begin{abstract}
Water is a crucial molecule in molecular astrophysics as it controls much of the gas/grain chemistry, including the formation and evolution of more complex organic molecules in ices. Pre-stellar cores provide the original reservoir of material from which future planetary systems are built, but few observational constraints exist on the formation of water and its partitioning between gas and ice in the densest cores. Thanks to the high sensitivity of the Herschel Space Observatory, we report on the first detection of water vapor at high spectral resolution toward a dense cloud on the verge of star formation, the pre-stellar core L1544. The line shows an inverse P-Cygni profile, characteristic of gravitational contraction. To reproduce the observations, water vapor has to be present in the cold and dense central few thousand AU of L1544, where species heavier than Helium are expected to freeze-out onto dust grains, and the ortho:para H$_2$ ratio has to be around 1:1 or larger. The observed amount of water vapor within the core (about 1.5$\times$10$^{-6}$\,M$_{\odot}$) can be maintained by Far-UV photons locally produced by the impact of galactic cosmic rays with H$_2$ molecules. Such FUV photons irradiate the icy mantles, liberating water wapor in the core center.  Our Herschel data, combined with radiative transfer and chemical/dynamical models,  shed light on the interplay between gas and solids in dense interstellar clouds and provide the first measurement of the water vapor abundance profile across the parent cloud of a future solar-type star and its potential planetary system.
\end{abstract}





\section{Introduction}

The ubiquity and importance of water in the Universe has become clear after the launch of the Herschel Space Observatory\footnotemark \footnotetext{Herschel is an European Space Agency space observatory with science instruments provided by European-led principal investigator consortia and with important participation from NASA.} \citep{pilbratt2010}. Water vapor has been found around moons and comets \citep{hartogh2011a, hartogh2011b}, in protoplanetary disks \citep{hogerheijde2011}, star forming regions \citep{ceccarelli10,vandishoeck2011}, starburst galaxies and active galactic nuclei \citep{vanderwerf2010, weiss2010}. 
One important class of objects is missing from this list: the centrally concentrated cores of molecular clouds, which precede the formation of stars and planetary systems, as previous studies have only provided upper limits of their water vapor content \citep{bergin02, klotz08}.   These so-called pre-stellar cores, as defined by \citet{crapsi05}, are centrally concentrated, with central densities $\geq 10^5$ cm$^{-3}$, they have central visual extinctions $\geq$50\,mag, large degrees of CO freeze-out ($\geq$ 90\%) and deuterium fraction ($\geq$20\%), and they show clear signs of contraction motions in high-density tracer spectra.  Pre-stellar cores are typically cold ($T <$ 20\,K) and detailed studies of two pre-stellar cores have found central temperatures of only 7\,K: L1544 \citep{crapsi07} and L183 \citep{pagani07}. 

Pre-stellar cores represent the initial conditions in the process of star and planet formation. Thus, a direct measurement of water toward a pre-stellar core would provide a unique opportunity to unveil the amount and the chemistry of water at the dawn of a new solar-type system.  Water has been clearly detected in solid form in molecular clouds, where pre-stellar cores are embedded, or in dense cores hosting young stellar objects \citep{whittet91, boogert08, chiar11}. However, these near- and mid-infrared solid-state observations cannot be extended to the realm of pre-stellar cores, given their large extinctions and starless nature. How can we measure the water abundance in a pre-stellar core?

The cold dust present in pre-stellar cores has its peak of emission at wavelengths $\geq$ 290\,$\mu$m and dust continuum emission is expected to be detectable at 557\,GHz, the frequency of the ground state transition of ortho-H$_2$O ($J_{K_A}-J_{K_C}$ = 1$_{10}$-1$_{01}$). Thus, the 557\,GHz water line (difficult to be collisionally excited in cold gas) could be seen in absorption against the far-infrared background produced by the core central regions.  \citet{caselli10} indeed detected an absorption feature at a 5\,$\sigma$ level using the  wide-band spectrometer (WBS) of the Heterodyne Instrument for the Far Infrared \citep[HIFI;][]{degraauw10} toward the pre-stellar core L1544. However,  the spectrum did not have enough sensitivity and spectral resolution to study the line profile and allow quantitative conclusions.   Here we report the results of a deeper search with 9 times greater spectral resolution and 2 times better sensitivity that has provided, for the first time, a velocity-resolved spectrum of water towards L1544, revealing the presence of water vapor in the central few thousand AU (the core center) for the first time.  The technical details of the observations are described in Section \ref{observations}, the data are presented in Section \ref{spectrum}, a closer look to the water line profile is given in Section \ref{inflow} and the data interpretation is described in Section \ref{model}.

\section{Observations and data processing}
\label{observations}

As part of the WISH (Water In Star forming regions with Herschel) key project \citep{vandishoeck2011}, the  ortho-H$_2$O(1$_{10}$-1$_{01}$) line \citep[frequency 556936.002$\pm$0.050\,MHz;][]{pickett05} was observed  with Herschel towards L1544 (coordinates $\alpha_{2000}$ = 05$^h$04$^m$17$^s$.21, $\delta_{2000}$ = 25$^{\circ}$10$^{\prime}$42$^{\prime\prime}$.8) for 3.3\,hours on March 20, 2010 and for 10.3\,hours on April 2, 2011. The data presented here are archived at the Herschel Science Archive\footnotemark \footnotetext{http://archives.esac.esa.int/hda/ui/}, under OBSID 1342192364 and 1342217688.  The half-power beam width (HPBW) of the telescope at this frequency is 40$^{\prime\prime}$. The technical setup in the two periods included dual beam switch mode, with the continuum optimization and fast chop options, and  simultaneous use of the wide-band (WBS) and the high-resolution (HRS) spectrometers in both H and V polarization. The individual spectra were reduced using the Herschel Interactive Processing Environment \citep[HIPE, version 6.1.0;][]{ott10}. Subsequent analysis of the data was performed with the Continuum and Line-Analysis Single dish Software (CLASS) within the GILDAS package\footnotemark \footnotetext{http://www.iram.fr/IRAMFR/GILDAS.}: the line intensity units were converted from antenna to brightness temperature ($T_{\rm B}$) using a beam efficiency of 0.76 \citep{roelfsema12}; the H and V spectra were analyzed individually and found similar within the noise (therefore, the two spectra were averaged). After binning to the nominal spectral resolution (0.59\,km\,s$^{-1}$ for WBS and 67\,m\,s$^{-1}$ for HRS), the root-mean-square (rms) noise level is 0.93\,mK in the WBS spectrum and 3.8\,mK in the HRS spectrum. The dust continuum emission flux at 557\,GHz (10.2$\pm$0.2\,mK) was obtained by dividing by 2 the intensity of the WBS 2011 dual beam switch spectrum, degrading its spectral resolution by a factor of 50 and measuring the mean and standard deviation of the brightness temperature of the remaining channels. 

\section{The water spectrum toward L1544}
\label{spectrum}

The (continuum-subtracted) spectrum of ortho-H$_2$O(1$_{10}$-1$_{01}$) observed with Herschel/HIFI toward L1544 is shown in Fig.\,1. In the top figure, the low spectral resolution HIFI-WBS spectrum of ortho-H$_2$O taken in 2010 \citep{caselli10} is superposed on the final (2010+2011) spectrum, confirming the absorption feature. The integrated intensity of the feature is now 10 times the associated error. The HRS spectrum smoothed at the WBS velocity resolution is also shown. The WBS absorption feature has a width at half-maximum of 2\,km\,s$^{-1}$, an optical depth of 1.0$\pm$0.2 and a column density of ortho-H$_2$O molecules in the ground state of (2.0$\pm$0.4)$\times$10$^{13}$\,cm$^{-2}$.  This is the first accurate water vapor column density estimate in a dark cloud, away from active sites of star formation. However, this water does not exclusively belong to L1544. In fact, the WBS absorption feature spans a large range of velocities and the bulk of the absorption appears shifted by about $+$2\,km\,s$^{-1}$ compared to the local standard of rest (LSR) velocity of L1544, $V_{\rm LSR}$ = 7.2\,km\,s$^{-1}$, marked by the dotted vertical line.  To unveil the water content in L1544,  velocity-resolved analysis is needed. 

The yellow strip in Fig.\,1a encloses the velocity range displayed in Fig.\,1b, where the final (2010+2011; red) and 2010 (black) HRS spectra of ortho-H$_2$O\,(1$_{10}$-1$_{01}$) are superposed on the CO ($J$ = 1$\rightarrow$ 0) spectrum \citep[grey histogram;][]{caselli10}. The intensities of the 2010 water line and the CO line have been reduced by a factor of 2 and 100, respectively, for displaying purposes. The final HRS spectrum reveals three features not noticeable in the upper panel: (1) an emission feature shifted by $-0.4\,{\rm km\,s^{-1}}$ compared to the $V_{\rm LSR}$ of L1544 (dotted vertical line); (2) an absorption feature centered at the $V_{\rm LSR}$ of L1544; (3) an absorption feature $+1.9$\,km\,s$^{-1}$ away from the $V_{\rm LSR}$ of L1544 and close to the secondary feature seen in CO(1-0), which is probably an independent cloud along the same line of sight (not to be considered further in this study as physical characteristics of this cloud are not yet known). The results of independent Gaussian fits to the three spectral components are listed in Tab.\,1. These features are spectrally diluted in the WBS spectrum, with the emission+absorption (components 1+2 in Fig.\,1b) toward L1544 resulting in the shallower absorption feature seen in the WBS spectrum at the velocity of  L1544.  The HRS spectrum thus provides information about the distribution of water along the line of sight.  

The water absorption feature seen at the velocity of L1544 can be used to estimate the amount of water vapor present {\it within} the pre-stellar core.  However, the intensity of the absorption is consistent with saturation (Tab.\,1), so only a lower limit can be gauged. From a curve-of-growth analysis, the optical depth of the L1544 absorption feature is $\tau > 1.5$. Assuming a H$_2$O ortho-to-para ratio of 3 \citep{lis10, herpin12} and a width at half-maximum of 0.5\,km\,s$^{-1}$ (Tab.\,1), the total (ortho+para) H$_2$O column density across the pre-stellar core is $>$1.0$\times$10$^{13}$\,cm$^{-2}$, corresponding to a fractional abundance (with respect to H$_2$ molecules) $>$1.4$\times$10$^{-10}$ \citep[using an H$_2$ column density $N({\rm H_2})$ $\simeq$ 7$\times$10$^{22}$\,cm$^{-2}$;][]{caselli10}. A detailed radiative transfer analysis is needed to unveil the amount of water within L1544 (see Section \ref{model}).

\section{Central water inflow}
\label{inflow}

To better understand the nature of the water emission in L1544, the water spectrum between 5.5 and 8.5\,km\,s$^{-1}$ is plotted in Fig.\,2 together with the ortho-H$_2$D$^+$(1$_{10}$-1$_{11}$) line \citep{caselli03}.  This line has a critical density of 10$^5$\,cm$^{-3}$ \citep{hugo09}, traces gas at radii of about 5000\,AU \citep{vastel06} and it is only moderately thick \citep{vandertak05}. The ortho-H$_2$O(1$_{10}$-1$_{01}$) line has a significantly larger critical density ($\geq$1.7$\times$10$^7$\,cm$^{-3}$, for an ortho-to-para H$_2$ ratio $\leq$ 3), and the line is highly subthermally excited in the region enclosed by the Herschel beam. These are ideal conditions to observe the characteristic ``infall asymmetry" \citep{evans99}: foreground cloud material, while collapsing, is absorbing the high-velocity portion of the line profile, producing
a brighter blue peak. In the presence of absorption against the continuum background emission, this is equivalent to an inverse P-Cygni profile, as observed in L1544. The water blue emission feature reveals the inner backside of the core, which is approaching the cloud center. Previous signatures of infall of pre-stellar cores were found using the N$_2$H$^+$(1-0) line \citep{crapsi05}, which traces gas with densities $\simeq$10$^5$\,cm$^{-3}$. The water line is tracing the gravitational contraction of gas at densities larger than 10$^6$\,cm$^{-3}$, which resides within a radius of 1000\,AU (see Sect. \ref{model}). Inverse P-Cygni profiles have also been found in 6 out of 15 Class 0 objects \citep{kristensen12}, highlighting water's ability to pick up infall motion along the line of sight.

The detection of emission (component \#1 in Fig.1b) is crucial for our understanding of where water resides in this system.  From chemical considerations alone it has been difficult to posit the presence of water in the dense center. We know that water ice is present in the less dense regions of the extended Taurus Molecular Cloud \citep{murakawa00}, accounting for almost 60\% of the total available gas phase Oxygen \citep[3.2$\times$10$^{-4}$ with respect to H nuclei;][]{meyer98}. Such a large amount of water cannot be produced via gas-phase routes and then deposited on top of dust grains, as gas-phase chemistry can only account for about one thousandth of the ice abundance \citep{lee96}.  Thus, it is now well-established that water ice is formed on the surface of dust particles via hydrogenation of atomic and molecular oxygen \citep{ioppolo08}. 

It is also well-established that atomic and molecular species start to freeze-out onto dust grains in pre-stellar cores at volume densities larger than a few times 10$^4$\,cm$^{-3}$ \citep{walmsley91, caselli99}.  L1544 is among the densest of these objects, with central values $\gtrsim$10$^6$\,cm$^{-3}$ (Sect. \ref{model}).  The relatively volatile CO molecules (evaporation temperature of $\sim$20~K) are mostly frozen onto grains in the core centers.  Water, with an evaporation temperature of $\sim 100$~K, should be more difficult to remove from dust grains and is therefore less likely to be observed via emission in the gas-phase. However, the detection of emission in our spectrum tells a different tale.  A small amount of water vapor must be present in the core center, contrary to our theoretical expectations, along with its presence in the outer layers seen in absorption (component \#2 in Fig.1b).

\section{The importance of cosmic-rays}
\label{model}

The chemistry, dynamical motions, and excitation are intimately intertwined in this object, resulting in saturated absorption and emission. This requires detailed physical, chemical and radiative transfer models to more accurately estimate the water content of L1544 and elucidate the key aspects of the velocity field in this object.  Current state-of-the-art chemical models \citep{hollenbach09} posit that small amounts of water vapor exist (containing less than a percent of the available cosmic oxygen) in dense regions that are well below the evaporation temperature, when exposed to the  far-ultraviolet (FUV) starlight pervading interstellar space.  When present,  UV photons photodesorb water molecules and other species from the icy mantles \citep{oberg09}.   Thus, the desorption depends on the intensity of the interstellar FUV field, which drops to tiny levels within dense clouds.   There is a small internal radiation field in these regions induced by cosmic rays interacting with H$_2$, leading to a dim FUV field about 1000 times less intense than the interstellar field \citep{gredel89}.  We have placed these mechanisms within the model of  \citet{hollenbach09}, which we follow here, in tandem with our dynamical \citep{keto08, keto10} and radiative transfer code MOLLIE \citep{keto90, keto04, keto10b}.   

Besides the comprehensive chemical network, we also consider a simplified oxygen chemistry to be easily coupled with our hydrodynamical code. The details of this chemical code will be presented in a separate paper (Keto et al., in preparation). Here we briefly mention that the formation of water is assumed to happen on dust grain surfaces by the reaction of OH and H. Gas phase routes to the formation of water in cold gas are not included. We allow gas-phase O and OH, as well as H$_2$O, to freeze-out onto dust grains, but assume that the frozen O and OH are immediately converted into H$_2$O ice. CO is treated independently from the oxygen chemistry. Photodesorption (including that due to the cosmic-ray induced FUV field), photodissociation and desorption due to direct cosmic-ray impacts with dust grains are also taken into account in the model, following \cite{hollenbach09} prescriptions. We adopt a cosmic-ray ionization rate of 1$\times$10$^{-17}$\,s$^{-1}$, as deduced by \cite{keto10}, and an average grain-surface cross section $\sigma_{\rm H}$=1.4$\times$10$^{-21}$\,cm$^{2}$, a factor of 1.4 lower than the value adopted by \citet{hollenbach09}. Overall, the difference between the simple and comprehensive model is within a factor of 2. 

Fig.\,3 (bottom panel) shows the physical structure of L1544, as deduced from a detailed comparison between observational data (N$_2$H$^+$(1-0), C$^{18}$O(1-0), C$^{17}$O(1-0) line profiles and the temperature structure deduced from NH$_3$) and simulated observations of a dynamical/chemical model of a spherical cloud near hydrostatic equilibrium undergoing gravitational contraction \citep{keto10}.  The middle panel displays three water vapor fractional abundance (with respect to H$_2$) profiles (color curves): (1) the red profile is the result of the full chemical network applied to the physical structure of L1544, without the internal FUV field produced by cosmic rays; (2) the green curve shows the water profile obtained in a self-consistent chemical-dynamical model using the simplified chemistry,  without the internal FUV field (note that results are similar to the more comprehensive model); (3) the blue thick curve is the result of the simplified chemistry with the inclusion of the dim FUV field produced by cosmic rays.   The black curve is the water ice abundance (divided by 10000) calculated with the full chemical network inclusive of the internal FUV field. The first thing to note is that the {\em FUV produced by cosmic-ray impacts is the main desorption agent in the core center}: water molecules are partially photodesorbed from the icy dust mantles and the water vapor abundance jumps by 3 orders of magnitude to $\geq$ 10$^{-9}$ within 5000\,AU. At larger radii, the visual extinction ($A_{\rm V}$) drops below 5\,magnitudes, allowing {\it interstellar} FUV photons to photodesorb icy mantles and to maintain the water abundance around 10$^{-7}$. Beyond 20000\,AU,  $A_{\rm V}$ becomes less than 2\,mag and the interstellar FUV photons efficiently photodissociate water, as well as other molecular species, explaining the drop in the water abundance. 

The simulated water lines obtained with the MOLLIE radiative transfer code, applied to the model cloud and water abundance profiles just described, are in the top panel of Fig.\,3. The MOLLIE code has been adapted to treat extremely sub-thermally excited lines \citep[such as the observed ortho-H$_2$O line; see also][]{snell00, poelman07}.  The blue, green and red line profiles refer to the corresponding H$_2$O abundance profiles in the central panel. The brighter blue (dashed) profile is obtained if  a 3:1 ortho:para ratio of H$_2$ molecules is assumed as collision partners of ortho-H$_2$O, while the fainter blue emission profile is obtained with a 1:1 ortho:para H$_2$ ratio \citep[using collisional coefficients from][]{dubernet09, daniel11}. Lower ortho:para ratios \citep[expected in regions with large amount of freeze-out;][]{pagani09,  troscompt09, parise11} are not consistent with the present observations, unless a higher abundance of water vapor is present in the gas phase at high densities.  Considering the simplicity of the physical  model, the agreement with the data is remarkable.  The similarity between the thick-blue curves and the data, including the continuum level,  confirms  that the observed water line proÞle is indeed the result of water molecules emitting in the core center and absorbing in the envelope. The observed extra-emission at low velocities suggests the presence of higher velocity motions, probably due to larger infall velocities within the central 1000\,AU than predicted by our models (Fig. \ref{h2o}).  This is the first tracer of dense gas known to be present near the core center that shows signs of gravitational contraction. 

In summary, Herschel has revealed the inflow of water vapor in a pre-stellar core and the key phenomena at work: (1)  fractional abundance of water vapor $\gtrsim$10$^{-9}$ toward the central few thousand AU (the core center). This is possible thanks to the cosmic-ray induced FUV field, which illuminates the icy dust mantles, allowing H$_2$O molecules to evaporate.  (2) Volume densities $\gtrsim$10$^6$\,cm$^{-3}$ in the core center. This allows ({\it i}) to collisionally excite the observed ground state rotational transition of ortho-H$_2$O and produce observable emission,  and ({\it ii}) to provide large enough dust continuum emission in the FIR to allow detection of absorption features.  (3) Gravitational contraction motions, which broaden the water line and allow the blueshifted emission to avoid the foreground ``water absorption screen". From the water vapor and ice abundance profiles deduced from our chemical models (Fig.\ref{h2o}), the total amount of water vapor within the central 10000\,AU is $\simeq$1.5$\times$10$^{-6}$\,M$_{\odot}$, while a reservoir of about 2.5$\times$10$^{-3}$\,M$_{\odot}$ (2.6 Jupiter masses) is available in the form of water ice.  

\acknowledgments
HIFI has been designed and built by a consortium of institutes and
university departments across Europe, the United States and Canada
under the leadership of SRON Netherlands Institute for Space Research.



{\it Facilities:} \facility{Herschel Space Observatory}

\clearpage




\clearpage

\begin{figure}
\includegraphics[angle=0,scale=.53]{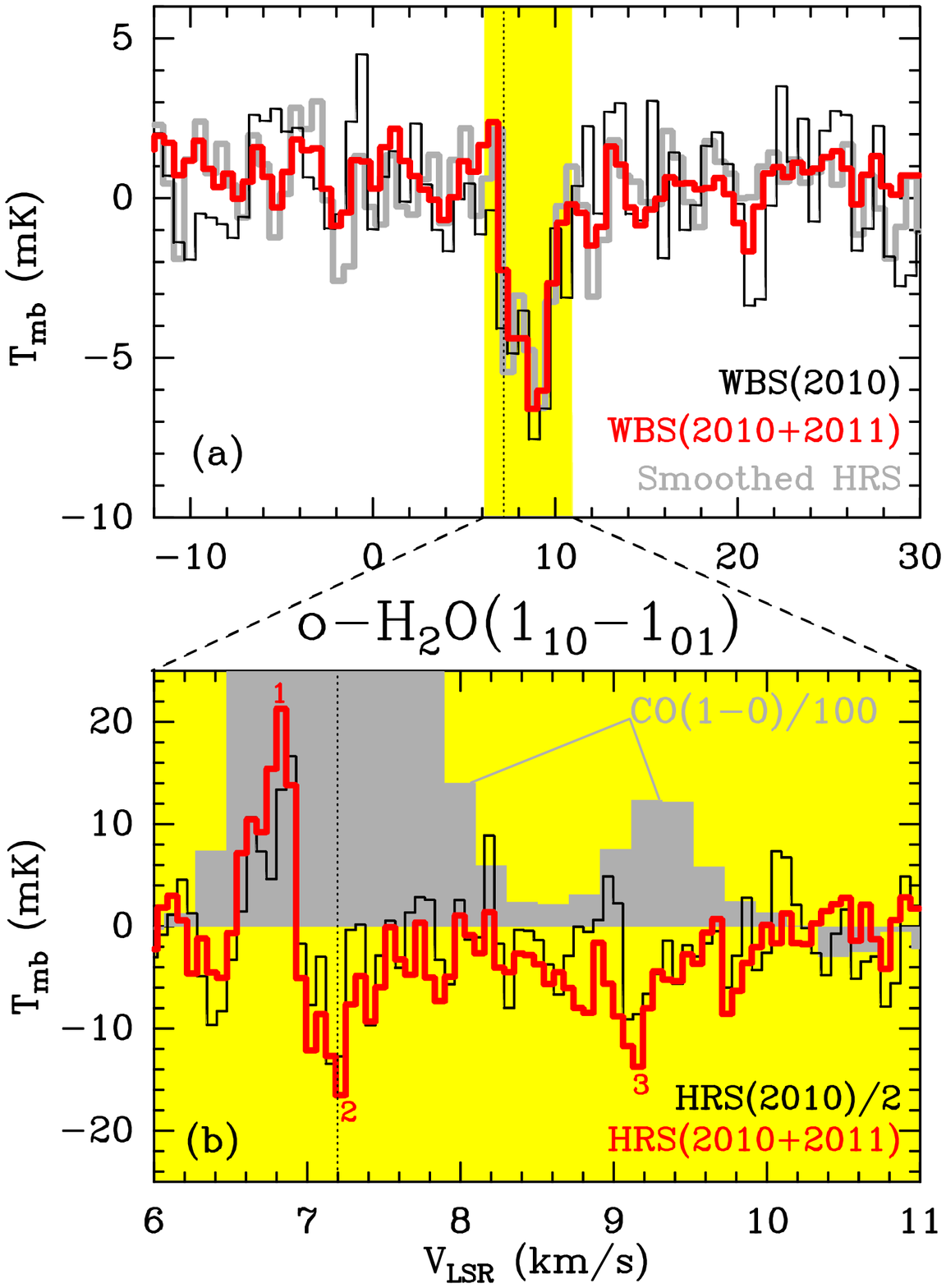}
\caption{{\bf (a)} The Herschel HIFI-WBS  spectrum (velocity resolution = 0.59\,km\,s$^{-1}$) of ortho-H$_2$O(1$_{10}$-1$_{01}$) observed toward L1544 in 2010 \citep[black histogram;][]{caselli10} and the final spectrum which combines  2010 and 2011 data (red histogram). The HRS spectrum smoothed at the velocity resolution of WBS is also shown (grey histogram). The dotted line marks the $V_{\rm LSR}$ of L1544 (7.2\,km\,s$^{-1}$).   {\bf (b)} HIFI-HRS spectrum (velocity resolution = 67\,m\,s$^{-1}$) of ortho-H$_2$O(1$_{10}$-1$_{01}$) (red: 2010+2011 data; black: 2010 data, intensity divided by 2). The grey histogram is the CO(1-0) line (intensity divided by 100), observed along the same line of sight \citep{caselli10}. Gaussian fit results and integrated intensities of the three labelled features are listed in Table 1. The continuum has been subtracted from all the spectra.} 
\label{DH}
\end{figure}

\clearpage

\begin{figure}
\includegraphics[angle=270,scale=.7]{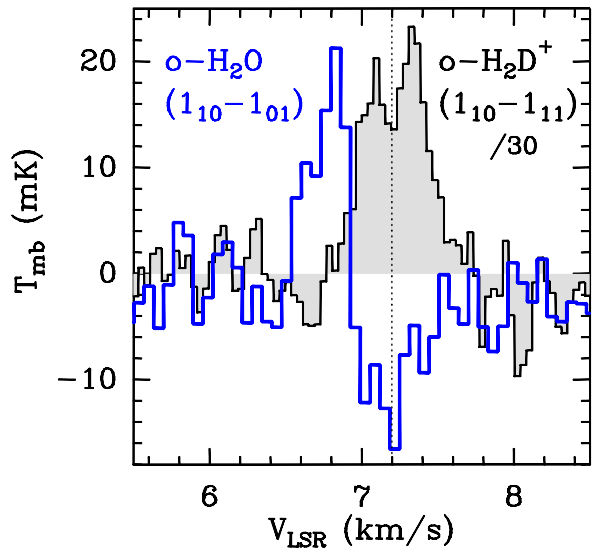}
\caption{HIFI-HRS spectrum of ortho-H$_2$O(1$_{10}$-1$_{01}$) (blue histogram) compared with ortho-H$_2$D$^+$(1$_{10}$-1$_{11}$) \citep[intensity divided by 30;][]{caselli03} toward L1544. The water line shows an inverse P-Cygni profile, with emission only at blueshifted velocities, whereas the water in the bulk of the core is absorbing the far-infrared dust continuum emission produced by the core central regions.}
\label{oco}
\end{figure}

\clearpage

\begin{figure}
\includegraphics[angle=0,scale=.53]{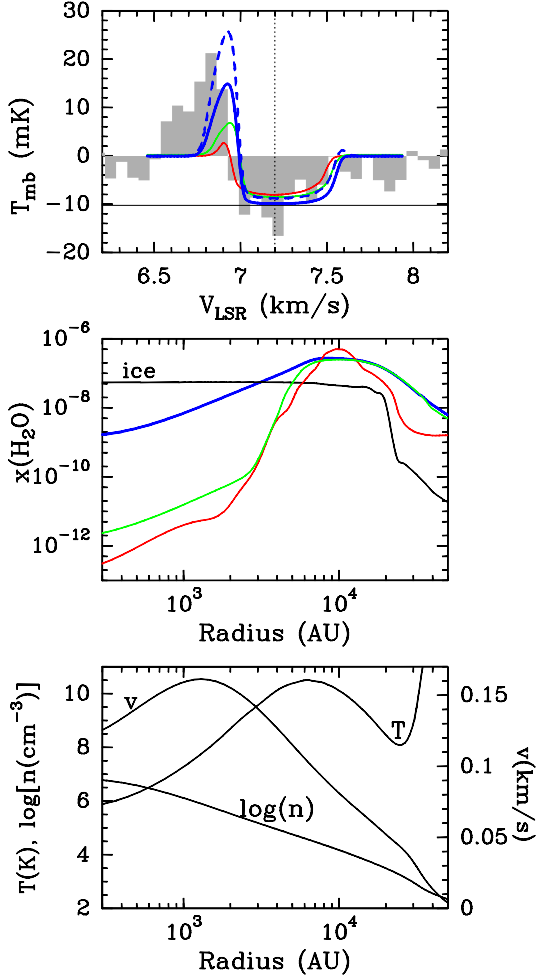}
\caption{{\it Bottom panel}: Physical structure of L1544 from \cite{keto10}: gas temperature, H$_2$ number density and velocity profiles. {\it Central panel}: fractional abundance of water vapor (w.r.t. H$_2$, color curves) as a function of core radius.  Red curve: comprehensive model without the cosmic-ray induced FUV field. Green curve:  simplified chemistry without the cosmic-ray FUV field.  Blue curve:  simplified model with the cosmic-ray FUV field. Black curve:  fractional abundance of water ice (divided by 10000) calculated with the full chemical code, inclusive of the cosmic-ray FUV field.  {\it Top panel}: ortho-H$_2$O(1$_{10}$-1$_{01}$) HRS continuum-subtracted spectrum (the horizontal black line displays the subtracted continuum level), compared with the results from the radiative transfer modeling, using as input the corresponding H$_2$O abundance profile in the central panel. The two blue curves refer to two different values of the H$_2$ ortho:para ratio, 3:1 (dashed) and 1:1 (thick).}
\label{h2o}
\end{figure}

\clearpage

\begin{table}
\caption{Gaussian fit results for the three features of ortho-H$_2$O(1$_{10}$-1$_{01}$). \label{gauss}}
\begin{tabular}{ccccc} \hline
Component & Area & $V_{\rm LSR}$ & $\Delta v$ & $T_{\rm B}$ \\ 
 & (mK\,km\,s$^{-1}$) & (km\,s$^{-1}$) & (km\,s$^{-1}$) & (K) \\ \hline 
1 & +4.7$\pm$0.8 & 6.79$\pm$0.02 & 0.24$\pm$0.05 & +19$\pm$4 \\ 
   & +5.0$\pm$0.7$^{a,1}$ \\
2 & -6.1$\pm$1.3 & 7.18$\pm$0.04 & 0.45$\pm$0.13 & -13$\pm$4 \\
   & -4.7$\pm$0.7$^{a,2}$ \\
3 & -9.4$\pm$1.7 & 9.05$\pm$0.10 & 1.12$\pm$ 0.25 & -8$\pm$4 \\
   & -6.8$\pm$1.1$^{a,3}$ \\ \hline 
\multicolumn{5}{l}{$^{(a)}$ Intensity integrated between channels with $|T_{\rm B}| > 4$\,mK:} \\
\multicolumn{5}{l}{$^{(1)}$ from 6.52 to 6.93\,km\,s$^{-1}$,} \\
\multicolumn{5}{l}{$^{(2)}$ from 6.98 to 7.45\,km\,s$^{-1}$,} \\
\multicolumn{5}{l}{$^{(3)}$ from 8.69 to 9.83\,km\,s$^{-1}$.} \\
\end{tabular}
\end{table}


\end{document}